\documentclass[a4paper,10pt]{article}

\usepackage[utf8]{inputenc}
\usepackage[T1]{fontenc}
\usepackage{amsmath,amsfonts,amssymb}
\usepackage{graphicx}
\usepackage{caption}
\usepackage{subcaption}
\usepackage{booktabs}
\usepackage{geometry}
\usepackage{ragged2e}
\usepackage[colorlinks=true, linkcolor=blue, citecolor=blue, urlcolor=blue]{hyperref}

\newcommand{\keywords}[1]{\par\noindent\textbf{Keywords: }#1}
\newcommand{\olsi}[1]{\,\overline{\!{#1}}}
\newcommand{\om}{\olsi{M}}
\newcommand{\oa}{\olsi{A}}

\title{A control theoretical approach to gene regulation raises quantitative constraints for dynamic homeostasis in stochastic gene expression}
\author{
    Guilherme Giovanini$^{a}$,
    Cyro von Zuben de Valega Negrão$^{b}$,
    Ammar Alsinai$^{c}$,\\
    Alexandre F. Ramos$^{a,d,1}$
    \\
    {\small\parbox{0.9\textwidth}{\justify
            $^a$Center for Translational Research in Oncology (LIM24)
                Comprehensive Center for Precision Oncology,
                Instituto do C\^{a}ncer do Estado de S\~{a}o Paulo (ICESP),
                Hospital das Cl\'{i}nicas da Faculdade de Medicina da Universidade de S\~{a}o Paulo (HCFMUSP),
                S\~{a}o Paulo,
                01246-000, SP, Brazil;
            $^b$Brazilian Biosciences National Laboratory (LNBio),
                Brazilian Center for Research in Energy \& Materials (CNPEM),
			   Campinas,
			   13083-970, SP, Brazil
		   $^c$Department of Mathematics,
		       CV Raman Global University,
		       Bhubansewar,
		       752054, Odisha, India
           $^d$Escola de Artes, Ci\^{e}ncias e Humanidades, Universidade de S\~{a}o Paulo,
                S\~{a}o Paulo,
                03828-000, SP, Brazil;
           $^{1}$To whom correspondence should be addressed. E-mail: alex.ramos@usp.br
     }}
}

\date{}

\begin{document}

\maketitle

\begin{abstract}
Cell phenotype dynamic homeostasis contrasts with the inherent randomness of intracellular reactions. Although feedback control of master regulatory genes (MRG) is a key strategy for maintaining gene network expression ranges limited, understanding the quantitative constraints and corresponding mechanisms enabling such a dynamic stability under noise remains elusive. Here we model MRG expression as a stochastic process and downstream genes as sensors which response conditionally induce MRG activity. We show that at homeostatic regime: {\it i.} the trajectories of the MRG expression levels can be adjusted towards specific ranges using both the exact solutions of the stochastic model and the exact stochastic simulation algorithm (SSA); {\it ii.} there exists a sampling rate which optimizes the feedback control of the MRG activity, and non-optimal controls resulting in alternative homeostatic dynamics; {\it iii.} the feedback control of MRG activity leads to updates which intensities and time intervals are non-linearly related; {\it iv.} the ON state probability of an MRG promoter has dynamics confined within a narrow domain. Our results help to understand the quantitative constraints underpinning dynamic homeostasis despite randomness, the mechanisms underlying alternative, non-optimal, homeostatic regimes, and may be useful for theoretically prototyping therapies aiming at gene networks modulation.
\end{abstract}

\keywords{Two-state stochastic gene, Master regulatory gene, Feedback control, Bursty gene expression}

\section{Introduction}\label{sec:intro}

Despite the major advances of molecular biology in the post-genomic era reconciling the randomness of intracellular processes underpinning dynamic homeostasis of cellular phenotypes remains elusive. The randomness is caused by chemical reactants being present inside the cells in low copy numbers \cite{Delbruck1940}. On the other hand, phenotypic homeostasis may be interpreted under a dynamical systems perspective as the ability of a system to have its trajectories within the neighborhood of a stable fixed point even under parameter value perturbations. Biologically, that corresponds to a cell reaching (and conserving) a given phenotype despite small variations of internal or external conditions. Noise, however, may eventually redirect a system towards a different stable fixed point, a deed corresponding to a cell reaching a new phenotype in the context of noisy variability of its internal conditions \cite{Guinn2020,Desai2021,Yang2022}. A mechanism to prevent those transitions is negative feedback control which provides phenotypic homeostasis \cite{Filo2023} and a strategy for regulating noise of biochemical processes and their cellular consequences \cite{Savageau1974,Becskei2000,Rosenfeld2002,Camas2006,Nevozhay2009,Ramos2015}. Frequently, the feedback control is implemented in a gene network whose dynamics is modulated by the expression levels of a master regulatory gene (MRG) and its multiple target genes \cite{Cai2020,Filo2023}. Since a plethora of systemic diseases, such as cancers, are associated with disruptions of a standard homeostatic profile of living systems towards alternative ones \cite{Kotas2015,Guinn2020}, understanding the mechanisms governing dynamical phenotypic stability is key for the design of more effective therapeutic strategies \cite{Thakore2016,Bulaklak2020,Yang2022}. One powerful approach is to design therapies as a control problem \cite{Jarrett2020} formulated to take the inherent stochasticity of the intracellular environment into consideration \cite{Giovanini2022}. That strategy may also be useful for understanding the mechanisms underlying cellular phenotypic homeostasis as driven by processes characterized by random fluctuations, which we do using a feedback control model for the modulation of the expression levels of a MRG.

\begin{figure}[!h]
\centering
\includegraphics[width=0.45\linewidth]{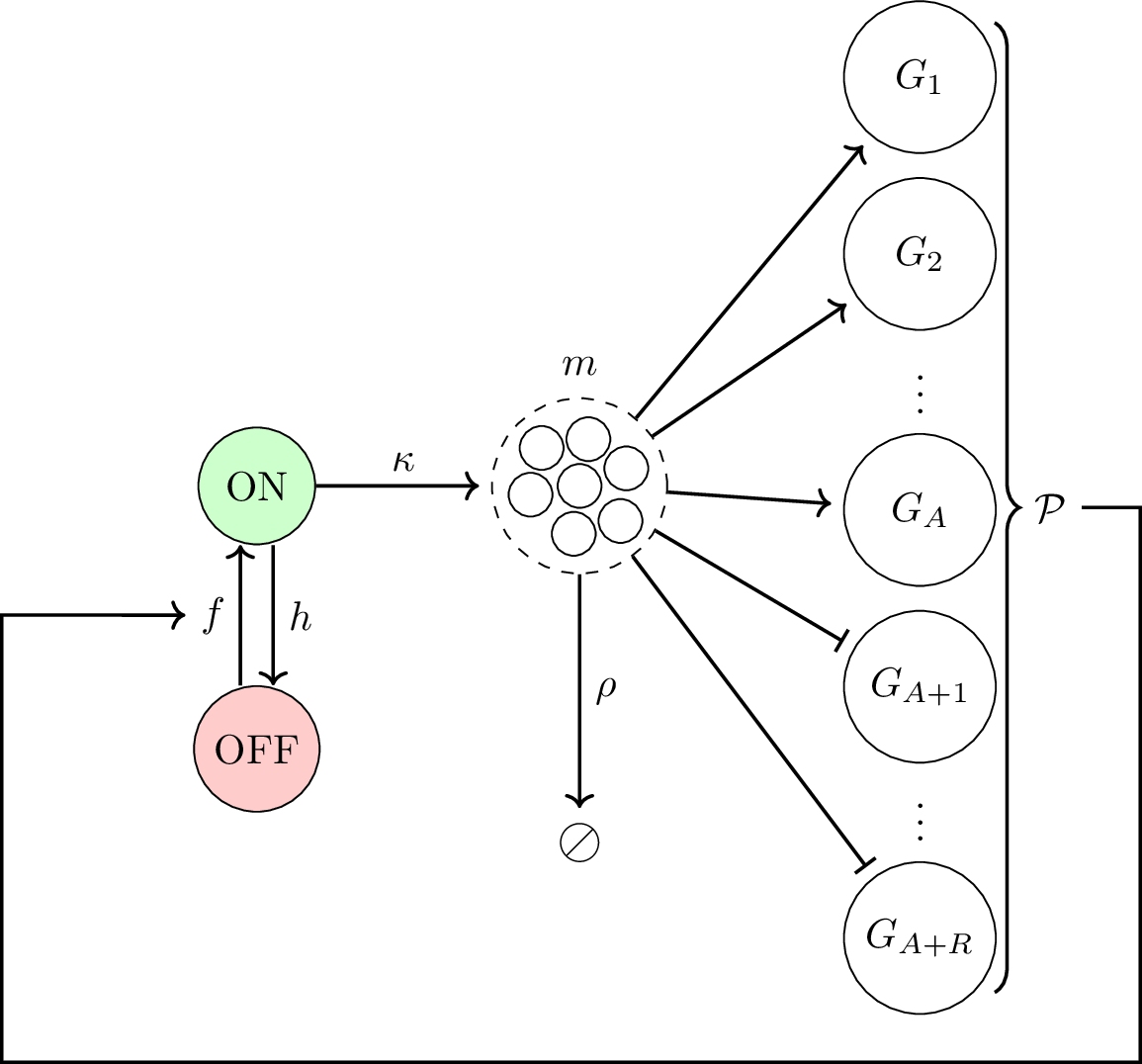}
\caption{{\bf A master regulatory gene (MRG) within a network gene.} The MRG expression synthesizes $m$ effective gene products over time. It has a promoter that switches between two states, ON and OFF. The number of MRG products increases  (or decreases) the expression levels of downstream genes ${\it G}_1, ...\,, {\it G}_A\, ({\rm or}\, {\it G}_{A+1}\,, ...\,, {\it G}_{A+R})$ The change of gene network state results in a feedback effect at MRG promoter state.} \label{fig:cartoon}
\end{figure}

In this study we combine a control theoretic approach and a two-state stochastic model for gene regulation \cite{Peccoud1995}, previously used to investigate bursts of gene expression in mouse and human fibroblasts \cite{Suter2011,Larsson2019}, to investigate the quantitative constraints and the mechanisms governing the homeostasis of a MRG. Fig. \ref{fig:cartoon} depicts the biological scenario considered by us. Our system describes the regulation of a MRG by a feedback effect dependent on the number of transcripts, denoted by $m$, expressed from a two-state MRG. The MRG RNAs are assumed to directly or indirectly induce an increase or decrease in the quantity of products of a set of target genes. The balance of the amounts of products from the target genes, as induced by the MRG RNAs, produce a $m$-dependent net effect which will loop back modulating the expression levels of the MRG. As an instance, let us consider the condition in which the number of RNAs from the MRG should be superior to $\om$. In that case, if $m < \om$ the activity of the MRG must be changed by means of some mechanism provided by the products of the target genes (here on denoted as feedback surge). We also assume that for $m \ge \om$ nothing happens. In this picture, it is implicit that the target genes operate as sensors, and the combination of products that they produce will be a readout of the state of the MRG as measured by the value of $m$. Since the products of the target genes are expected to be produced in bursts \cite{Suter2011,Larsson2019}, we implement an effective sampling of $m$ to mimic the interaction between the RNAs and their target genes. Using this model we show in this manuscript: {\it i.} that the trajectories of the amounts of RNAs from the MRG can be adjusted towards specific values using either the exact solutions for $\langle m \rangle (t)$ obtained from the two-state model \cite{Prata2016} or the algorithm for exact stochastic simulations of chemical reactions \cite{Gillespie1976}; {\it ii.} the existence of a sampling rate which optimizes the feedback surges controlling the activity of the MRG; {\it iii.} the feedback surges generated by net effect have a nonlinear relationship between the intensities and time intervals and; {\it iv.} the dynamics of probability for the ON state of MRG promoter is confined within a narrow domain during homeostatic regime. The presented model provides insights into how homeostasis emerges despite randomness, indicates that rebalancing the time scales of the sensor system enables redirecting the dynamic homeostatic regime of a MRG. We consider that our approach may be used as a prototyping tool for the design of therapies aiming to modulate gene networks.

In Sec. \ref{sec:methods}, we describe the theoretical formulation for the coupling between the stochastic binary MRG model and the transcripts-dependent feedback control of MRG expression by the gene network effect. Sec. \ref{sec:results} presents the results about the control of homeostatic MRG expression, characterizing the emergent behavior of MRG variables over the range of parameter values. A discussion about our results, limitations of our approach, and open perspectives is presented in Sec. \ref{sec:discussions}.

\section{Methods}\label{sec:methods}

\noindent{\bf Phenomenological coupling of feedback control to a two-state stochastic model for regulation of gene transcription.}
We model a MRG using the exactly solvable two-state stochastic model \cite{Peccoud1995,Biswas2009}, which is widely used as a basic building block to understand noise in gene expression \cite{Innocentini2007,Shahrezaei2008,Ramos2010,Prata2016,Gama2020,Suter2011,Zoller2018,Silva2025}. The model is depicted by the scheme shown in Fig. \ref{fig:cartoon}. It has two random variables $(s,m)$, with $s\in\{\mathrm{ON}, \mathrm{OFF}\}$ being the state of the promoter, and $m\in\{0,1,2,\ldots\}$ denoting the number of RNAs. The promoter randomly switches from state OFF to ON (and vice-versa) with a rate $f$ (and $h$). The synthesis and degradation rates are respectively denoted by $\kappa$ and $\rho$, with synthesis only happening when the promoter is ON. We also introduce a modification: the number of transcripts is monitored with a sampling rate $\nu$, and if the number of transcripts is lower than a pre-set aimed value, one or more rates governing the state of the system should be changed to drive expression back to the aimed level $\om$. Here we choose to only increment the value of $f$, as that choice provides sufficient intricacy for clearly demonstrating our theoretical formulation.

The aforementioned effective chemical reactions are summarized below. Let us denote by: $\mathcal{R}$, the promoter of the gene; $\mathcal{P}$, the balance of products from the sensing genes producing a net effect which may (or may not) induce an increase on OFF to ON rate ($f$) of the MRG; and $\Delta\mathit{RNA}$, the aimed change on the number of RNAs to restore its expression level towards its aimed value $\om$. The latter transition aims to rebalance $\mathcal{P}$ by an amount $\Delta\mathcal{P}$. That sets a gene network functioning as a feedback system incrementing $f$ as indicated by the balance on the number of products synthesized from the target genes. Functional transcripts of MRG are represented by {\it RNA} and degradation of, or loss of functionality by, RNAs is indicated by $\oslash$:
\begin{eqnarray}
	\mathcal{R} + \mathcal{P}&
	\stackrel{f(\mathcal{P})}{\rightharpoonup}&
	\mathcal{RP}, \label{eq:ecr1} \\
	\mathcal{RP}&
	\stackrel{h}{\rightharpoonup}&
	\mathcal{R} + \mathcal{P}, \label{eq:ecr2} \\
	\mathcal{RP}&
	\stackrel{\kappa}{\rightharpoonup}&
	\mathcal{RP} + \mathit{RNA}, \label{eq:ecr3} \\  
	\mathit{RNA}&
	\stackrel{\rho}{\rightharpoonup}&
	\oslash, \label{eq:ecr4} \\
	\mathcal{P}&
	\stackrel{\nu}{\rightharpoonup}&
	\mathcal{P} + \Delta\mathcal{P}(\Delta\mathit{RNA}). \label{eq:ecr5}
\end{eqnarray}
Equations~\ref{eq:ecr1} -- \ref{eq:ecr5} respectively indicate promoter switching from OFF to ON, and ON to OFF, synthesis and degradation of RNAs, and the sampling of RNAs number. The rate $f \equiv f(\mathcal{P})$ indicates that a differential configuration on the number of products from the target genes will affect the OFF to ON switching rate (Fig. \ref{fig:cartoon}).

Here we consider the time-dependent solutions of the moments of the probability distributions governing~$(s,m)$~\cite{Prata2016}. Let us set the  parameters $\epsilon$, $A_s$ and $N$ respectively denoted as the ratio of the gene switching rate between ON and OFF states and the degradation rate of the gene products, the steady state probability for the promoter to be ON, and the steady state average number of products for a promoter fully ON:
\begin{equation}\label{eq:parms}
	\epsilon=\frac{f + h}{\rho}, \ \ \
	A_s=\frac{f}{f + h}, \ \ \
	N=\frac{\kappa}{\rho}.
\end{equation}
For parameter values being constant, one may write the equations governing the dynamics of the probability for the promoter being ON, $A(t)$, and the average number of products, $\langle m \rangle (t)$, as
\begin{eqnarray}
	A(t) &=&
	A_s +
	(A_0 - A_s)\,\mathrm{e}^{-\epsilon \rho t}, \label{eq:probON_t} \\
	\langle m \rangle (t) &=& \langle m \rangle_s +
	Y\,{{\rm e}^{- \epsilon \rho t}} +
	V \, {{\rm e}^{- \rho t}},\label{eq:avrg}
\end{eqnarray}
where $A_0=A(0)$, and $\langle m \rangle_0=\langle m \rangle(0)$ are initial conditions, $\langle m \rangle_s = N A_s$ is the steady state average number of gene products, $Y = N\dfrac{A_0 - A_s}{1 - \epsilon}$, and $V = \langle m \rangle_0 - \langle m \rangle_s - Y$.

Because the feedback leads to a time-dependence on the kinetic parameters of the model, one must adapt the above solutions. Recently, we proposed a piecewise decomposition of the dynamical regime of the system \cite{Giovanini2022}. The exact solutions for constant parameter values are used within each time interval when they are sufficiently small with the state at the end of one interval being the initial condition of the next one. That enabled us to use the solutions presented at Eqs.~\ref{eq:probON_t} and \ref{eq:avrg}
\\

\noindent{\bf An approach for investigating homeostatic gene expression by feedback-based modulation of the OFF to ON switching rate.}
To illustrate our methodology, we consider a gene whose expression level in a normally functioning cell must be high. The parameter values are set to drive a steady state regime having low expression levels when the feedback system is not operational. That condition corresponds to a MRG being strongly repressed and in a cell operating in an abnormal regime. Hence, as an initial condition we assume the MRG being in a low expression level steady state regime set in terms of a small OFF to ON switching rate $f_s$. Hence, the feedback surges will operate to increase $f$ and, consequently, the number of transcripts (see Eq. \ref{eq:ecr1}). We assume that the effect of a feedback surge on $f$ decays exponentially with rate $\lambda$, and that $f$ returns to $f_s$ \cite{Giovanini2022}. Therefore, once the feedback surges start, $f$ becomes time-dependent, $f_t$. After a fixed time interval $\Delta t$ determined by sampling rate $\nu$, the need to increase $f$ is verified by checking the number of transcripts. We assume that the change in $f$ is sufficiently fast to be approximated as instantaneous such that the update in $f$, when needed at $t + \Delta t$, is given by:
\begin{equation}\label{eq:dyn_f}
	f_{t + \Delta t} =  f_s + (f_t - f_s)\,{\rm e}^{-\lambda\,\Delta t} + \Delta f_t,
\end{equation}
where $\Delta f_t$ is a positive number set accordingly with the difference between $\langle m \rangle (t)$ and $\om$.
\\

\noindent{\bf A phenomenological proposal for updating $f$.} 
As a first order approximation for proposing a feedback control, we consider $\langle m \rangle (t)$ to be in a hypothetical steady state value at instant $t$ and $t + \Delta t$. Hence, this implies on assuming the average number of RNAs being approximated by the hypothetical steady state values $f = f_t$ and $f = f_{t+\Delta t}$, such that the corresponding averages, $M_t$ and $M_{t+\Delta t}$, are
\begin{eqnarray}
	M_t &=& \frac{\kappa}{\rho}\frac{f_t}{f_t + h} \label{eq:mean_s_t}, \\
	M_{t+\Delta t} &=& \frac{\kappa}{\rho}\frac{f_t + \Delta f_t}{f_t+h + \Delta f_t} = \om. \label{eq:mean_s_dt}
\end{eqnarray}
The feedback increment $\Delta f_t$ is obtained from the difference $\Delta M_t = M_{t+\Delta t} - M_t = \om - M_t$ which is a positive quantity as $M_{t+\Delta t}$ always set to be equals to $\om$. Then, the OFF to ON switching rate in Eq.~\ref{eq:dyn_f} is changed by:
\begin{equation}\label{eq:f_update}
        \Delta f_t = \dfrac{h+f_t}{N-\om} \, \Delta M_t,
\end{equation}
which is always positive because $N>\om$ by construction. The inverse function of the Eq.~\ref{eq:f_update} has a sigmoidal shape on $f_t$. Hence, $\Delta M_t$ is limited by a threshold value even for large $\Delta f_t$.
\\

\noindent{\bf The stochastic simulation algorithm (SSA),} or Gillespie algorithm \cite{Gillespie1977}, is a rigorous approach for performing exact simulations of the evolution of systems of chemical reactions. The dynamics produced by the effective chemical reactions shown in Eqs. \ref{eq:ecr1}--\ref{eq:ecr5} were obtained using the SSA. Equation~\ref{eq:ecr5} indicates the reaction that monitors the number of gene products to produce the increments in $f$ by means of a feedback. That is the stochastic analog of the procedure adopted to insert a feedback in the averages as shown in Eq.~\ref{eq:dyn_f}. In the SSA scheme, we also assume an exponential decay and random instantaneous increments of $f$, and $\nu$ as the propensity of occurrence of the reaction \ref{eq:ecr5}.
\\

\noindent{\bf Dynamics of the ON state probability and average number of transcripts as a control system.}
As mentioned above, Eqs. \ref{eq:probON_t} and \ref{eq:avrg} are solutions of the following ODE system \cite{Giovanini2022}:
\begin{eqnarray}
  \dfrac{dA}{dt} &=&
  -\left(f+h\right)A
  +f,\nonumber
  \\
  \dfrac{d\langle m\rangle}{dt} &=&
  -\rho\langle m\rangle
  +\kappa A, \label{eq:ODE_system}
\end{eqnarray}
where we omitted the time-dependence of $f$, $A$ and $\langle m \rangle$. The steady-state $A_s$ and $\langle m \rangle_s$ in Eqs.~\ref{eq:parms} and~\ref{eq:avrg} are computed by equating the left-hand side of the ODE system to 0. This ODE system is a linear control system, $\dot{{\bf x}}(t) = {\cal A}(t){\bf x}(t) + {\cal B}(t){\bf u}(t)$, where the state vector ${\bf x}(t)$ and input ${\bf u}(t)$ are respectively:
\begin{equation}
{\bf x}(t) =
\begin{bmatrix}
A(t) - A_s\\
\langle m \rangle(t) - \langle m \rangle_s
\end{bmatrix}
, \ \
{\bf u}(t)=f_t.
\end{equation}
The matrices $({\cal A}$, ${\cal B})$ represent the homogeneous and non-homogeneous components of the system, and using Eqs.~\ref{eq:ODE_system} we obtain:
\begin{equation}
{\cal A}(t) =
\begin{bmatrix}
-(f_t+h) & 0 \\
\kappa & -\rho
\end{bmatrix}
, \ \
{\cal B} =
\begin{bmatrix}
1\\
0
\end{bmatrix}
\,. \label{eq:matrices}
\end{equation}
Note that ${\cal B}$ enables the control to be established.

The state of the system at $t$, ${\bf x}(t)$, is obtained by the action of a transition matrix $\Phi_{\cal A}(t,t_0)$ related to the matrix ${\cal A}$ on ${\bf x}(t_0)$. The solution ${\bf x}(t)$, which exists and is unique, can be computed by means of the celebrated Dyson series (as known as Peano-Baker series) \cite{Dyson1949,Rugh1996}. The closed form of that solution for time-dependent rates is beyond the scope of the current study. Because we are investigating the phenomenology underlying the homeostasis of cellular phenotype using the simplest possible theory for regulation of stochastic gene expression, a numerical analysis of the solutions suffices.

Notice, though, that we do have the closed forms of the solution for constant kinetic parameters which will be useful in our numerical computations. Let us partition the time domain in $I$ subintervals, namely $[t_0,t]=[t_0,t_1)\cup[t_1,t_2)\cup\dots[t_{I-1},t_I]$. During each $\Delta t_i=t_i-t_{i-1}$ we may assume that all kinetic parameters are constants within an arbitrarily defined precision. Because we presume an exponential decay of the effect of the control onto a given kinetic rate, the length of the subintervals vary (see Ref. \cite{Giovanini2022} for a description). During each subinterval $[t_{i-1},t_i)$, the transition matrix is:
\begin{equation}
\Phi_{\cal A}(t_{i},t_{i-1}) =
\begin{bmatrix}
{\rm e}^{-(f+h)\Delta t_i} &
0
\\
\frac{\kappa \big({\rm e}^{-(f+h)\Delta t_i} - {\rm e}^{-\rho\,\Delta t_i}\big)}{\rho\, - (f+h)} &
{\rm e}^{-\rho\,\Delta t_i}
\end{bmatrix}
\,. \label{eq:transition_matrix}
\end{equation}
The composition property of transition matrix, $\Phi_{\cal A}(t_{i},t_0) = \Phi_{\cal A}(t_{i},t_{i-1})\Phi_{\cal A}(t_{i-1},t_0)$, enables the piecewise approach to perform small parameter variation within ${\cal A}$.

The controllability of the system is assessed by means of the eigenvalues of ${\cal A}$. This matrix has two eigenvalues $-\rho$ and $ -(f + h)$ with
\begin{equation}
  \left[
  \begin{array}{c}
    0 \\
    \frac{f+h}{f+h-\rho}\left(\langle m \rangle(t_{i-1}) - \langle m \rangle_{s}\right)
  \end{array}
  \right]
  \ \ \ \ 
  \left[
    \begin{array}{c}
      {\rm A}{(t_{i-1})} - A_{s} \\
      \frac{\rho}{\rho - f - h}\left(\langle m \rangle(t_{i-1}) - \langle m \rangle_{s}\right)
    \end{array}
    \right]
\end{equation}
being their respective eigenvectors. Since $\rho, f, h$ are positive real, all eigenvalues are negative real ensuring that the system is exponentially stable (see Theorem 6.10 in \cite{Rugh1996}). The controllability matrix $[{\cal B} \quad {\cal AB}]$ has full rank and, hence, the system is controllable (Theorem 9.5 \cite{Rugh1996}). Considering these theoretical results for the time-invariant system, we will act onto $f$ parameter to investigate the stability and controllability of ${\bf x}(t)$.

\section{Results}\label{sec:results}
We simulate the trajectories of the number of transcripts starting from an initial condition as $\langle m \rangle_{s}=10$. The feedback surges were set to cause a 10-fold increase on the number of gene products, {\it i.e.} from 10 to 100. The sampling frequencies, $\nu$, underpin the surge of the feedback span from $1 \times 10^{-2}$ to $2 \times 10^{2}\rho$. For simplicity we set $\rho= 1\,{\rm TU}^{-1}$, where TU denotes the time unit corresponding to the half-life of the transcripts which is used to set the time scale of our system. The values of the kinetic rates $(f_s, h, \kappa)$ are $(0.9,\, 9.1,\, 110)$ in units of ${\rm TU}^{-1}$. These parameters characterize a quasi-Poissonian probability distribution governing the RNA number in the steady-state regime \cite{Giovanini2020, Giovanini2022}. Using Eqs.~\ref{eq:parms}, the distributions can also be characterized by another auxiliary set of phenomenologically interpretable parameters: $(\epsilon, A_s, N) = (10,\, 0.09,\, 110)$. Before the beginning of the feedback surges, we consider that the system is in a steady-state regime. The first feedback surge occurs at instant 1 TU, and we follow the dynamics of the system until $t=10^3$ {\rm TU}. The intensity of the effect of the first increment is 90 for all trajectories because of the aimed 10-fold increase in $\langle m \rangle$. To investigate how the decaying rates of the effects of the feedback surges affect the control, we use the following values of $\lambda$: $(0.01\,, 0.1\,, 0.5\,, 1\,, 2)$. The time steps of the dynamics are computed using a piecewise approach applied to the exponential decay function $f(t)$, as described in \cite{Giovanini2022}, where the absolute error of each subinterval and the stopping criterion, $f(t)-f_s$, are both $1 \times 10^{-8}$.

\subsection{Sampling rate enables regulation of homeostatic RNA levels}

\begin{figure}[!h]
\includegraphics[width=1\linewidth]{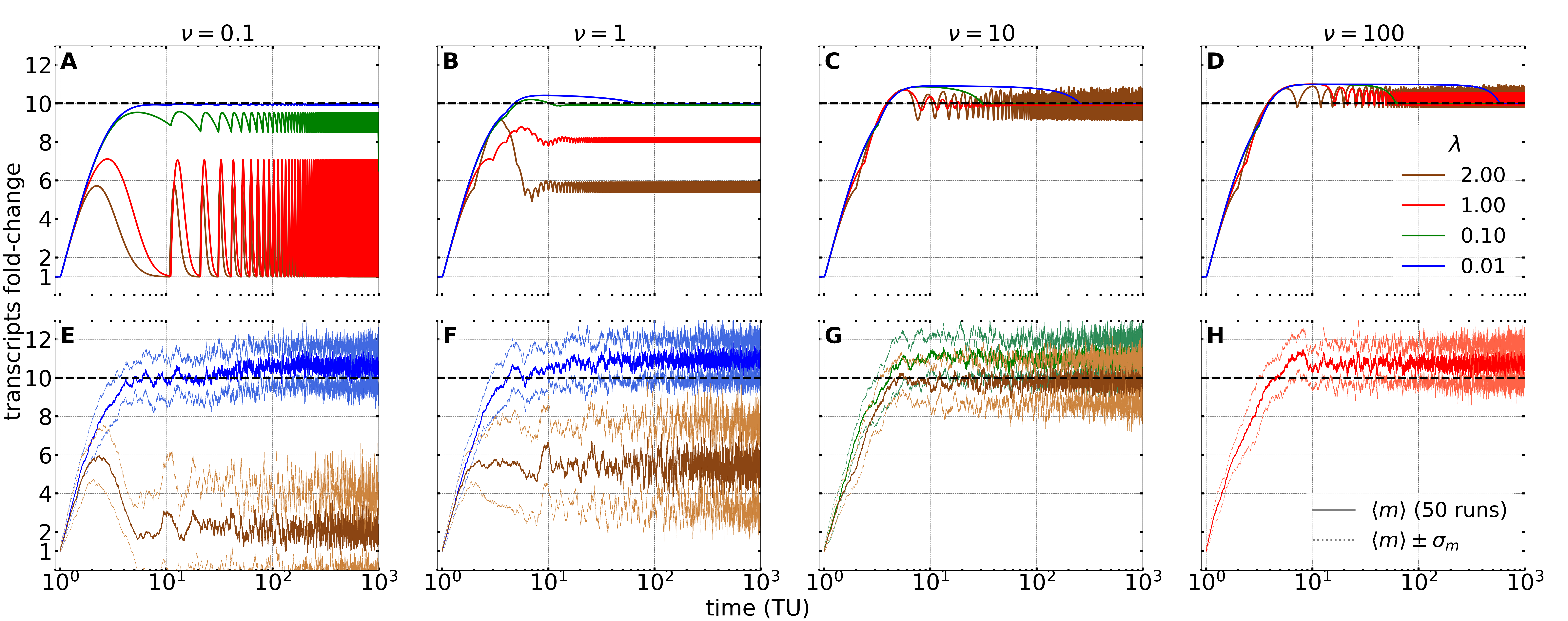}
\caption{{\bf Sampling rate regulates RNA levels at homeostasis.} From left to right, the columns present trajectories with increasing sampling rate $\nu$. All curves follow the color key within {\bf (D)} for four different decaying rates of increment effect $\lambda$. {\bf (A--D)} depict analytical trajectories of the fold-change in the average RNA levels $\langle m \rangle$. {\bf (E--H)} show average and standard deviation of RNA numbers computed by 50 runs using the Gillespie algorithm. A dashed line at 10 indicates the fold-change corresponding to $\om$. The scales of the rates and time are relative to the RNA degradation rate $\rho$.} \label{fig:fold_m_cumulative_vs_t}
\end{figure}

Figure~\ref{fig:fold_m_cumulative_vs_t} shows the dynamics of the fold-change of the average numbers of transcripts computed using either Eq.~\ref{eq:avrg} (top row) and fifty trajectories obtained by the SSA simulation of the reaction scheme of Eqs.~\ref{eq:ecr1}--\ref{eq:ecr5} (bottom row). We used four sampling rate values (see columns) and four decaying rates of the feedback effect (color key in graph {\bf (D)}). For a sampling at time $t$, the amount of RNAs $\langle m \rangle(t)$ is compared to the aimed value $\om$. For $M_t<\om$ the feedback surges and $f_t$ is incremented by $\Delta f_t \propto \Delta M_t$ to induce an increase in the number of transcripts.

The increase of $\nu$ stabilizes $\langle m \rangle_t$ around $\om$ whether we use the analytically obtained, or SSA dynamics, typically with reduced variability. In graph {\bf (A)}, the average number of transcripts reaches homeostasis for $\lambda = 0.01$. For $\nu \geq 1$, graphs {\bf (B--D)}, the trajectories for $\lambda < 1$ exhibit dynamics that overshoot and then dampen towards the $\om$. For $\lambda \geq 1$, the homeostatic regime shows an oscillatory-like behavior. As $\nu$ increases, the oscillations surround $\om$, with more heterogeneous amplitudes and a larger band. Lower sampling rates, $\nu = 0.1$, do not allow the $\langle m \rangle_t$ dynamics to stabilize when $\lambda \geq 0.1$. Note, however, the existence of trajectories exhibiting a dynamic homeostasis around intermediary values of $\langle m \rangle$ which are regulated by the relation between $\nu$ and $\lambda$. This behavior is also observed in the trajectories obtained using Gillespie algorithm.

Comparing the analytical curves with SSA ones, it is noticeable that for $\nu = 0.1$, {\bf (A)} shows fast increases on the amount of transcripts, while in {\bf (E)}, the curve with larger $\lambda$ shows less controllable behavior (brown curve). When $\nu = 1$, the peak that appears around $t = 3$ TU in {\bf (B)} for $\lambda = 2$ (brown curve) does not occur in graph {\bf (F)}, because only a few trajectories obtained by SSA tend to raise at this instant. For $\nu \geq 10$, the amplitudes of the bumps in the number of transcripts become more similar in the analytical and SSA curves. Because the SSA trajectories behave similarly to those obtained from the analytical solutions, we use the analytical solutions to analyze the properties of the feedback control proposed by us. Note, however, that our theoretical analysis will show results that can be extrapolated for the realization of the stochastic process as simulated by the SSA and, consequently, on the analysis of experimental data.

\subsection{Sampling rate reveals an optimum for the average feedback surge effects}

\begin{figure}[!h]
\centering
\includegraphics[width=1\linewidth]{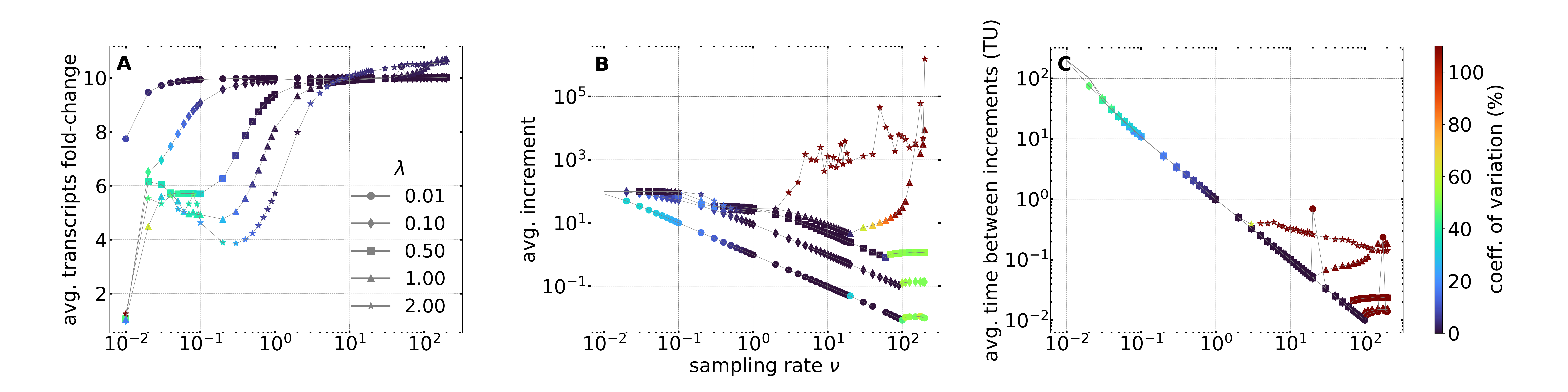}
\caption{{\bf Optimal sampling rates minimizing average feedback surges intensities.} The graphs depict the dependency of three variables on the sampling rate: the average fold-change in RNA levels {\bf (A)}, the average intensity {\bf (B)}, and the time interval {\bf (C)} of the increment effect for dynamics at the homeostatic regime. Marker colors indicate the coefficient of variation (CV) in percentage for each respective y-axis variable. Rate and time scales are relative to the RNA degradation rate $\rho$. All x-axes and the {\bf (B--C)} y-axes are in logarithmic scale.} \label{fig:m_fold_dosef_tauf_vs_fsampling_bar_cv}
\end{figure}

Figure~\ref{fig:m_fold_dosef_tauf_vs_fsampling_bar_cv} has the sampling rate $\nu$ at horizontal axes while the vertical axes of graphs {\bf (A)},  {\bf (B)}, and  {\bf (C)}, respectively present, at the homeostatic regime, the average fold change of $\langle m \rangle$, the intensities of increments of feedback surges and time intervals between these increments. The homeostatic regime is defined for $t \geq 900$ TU. This interval ensures that for larger $\nu$, all $\langle m \rangle$ trajectories fluctuate within a defined band approaching the aimed level. The color code on the right indicates coefficient of variation (CV) values of the variables of the corresponding vertical axis. Each symbol style indicates a different value of $\lambda$ as shown in graph {\bf (A)}.

Graph {\bf (A)} shows that $\langle m \rangle$ reaches $\om$ for larger $\nu$. It is noteworthy that $\langle m \rangle$ exceeds $\om$ for $\nu > 50$ when $\lambda = 1$, and for $\nu \geq 10$ when $\lambda = 2$. For $\nu < 10$, $\langle m \rangle$ decreases and its CV increases for all $\lambda$. Local minima of $\langle m \rangle$ fold-change are observed for $\nu$ between $0.04$ and $0.3$. The minimum at the lowest $\nu$ occurs for $\lambda = 0.5$ with $\langle m \rangle$ approximately 5.5-fold; and the one at the highest $\nu$ occurs for $\lambda = 2$ with $\langle m \rangle$ around 4-fold. For $\lambda \leq 0.1$, $\langle m \rangle$ decreases monotonically as $\nu$ decreases.

Graph {\bf (B)} shows the decrease in average increment as $\nu$ increases until a $\lambda$ dependent threshold. From the lowest to the highest $\lambda$, the minimum average increment occurs for $\nu = (100,\, 90,\, 60,\, 20,\, 2)$. The CV of increment effects increases for $\nu$ higher than that minimum one, and average increment becomes higher and unstable for $\lambda \geq 1$. In this case, increments surpass those of lower $\nu$.

In graph {\bf (C)}, as $\nu$ increases, the average intervals between increments form a descending straight line until they reach a threshold. From the smallest to the second-largest $\lambda$, the respective $\nu$ that minimizes the average time intervals are $(100,\, 100,\, 60,\, 20)$. Note that for $\lambda = 2$, no minimum occurs, and in this case, the time intervals decrease slowly for $\nu > 3$. Similar to {\bf (B)}, if $\nu$ is larger than the threshold, the CV of the increment interval increases.

\subsection{A nonlinear relationship between time intervals and intensities of feedback surges}

\begin{figure}[!h]
\centering
\includegraphics[width=1\linewidth]{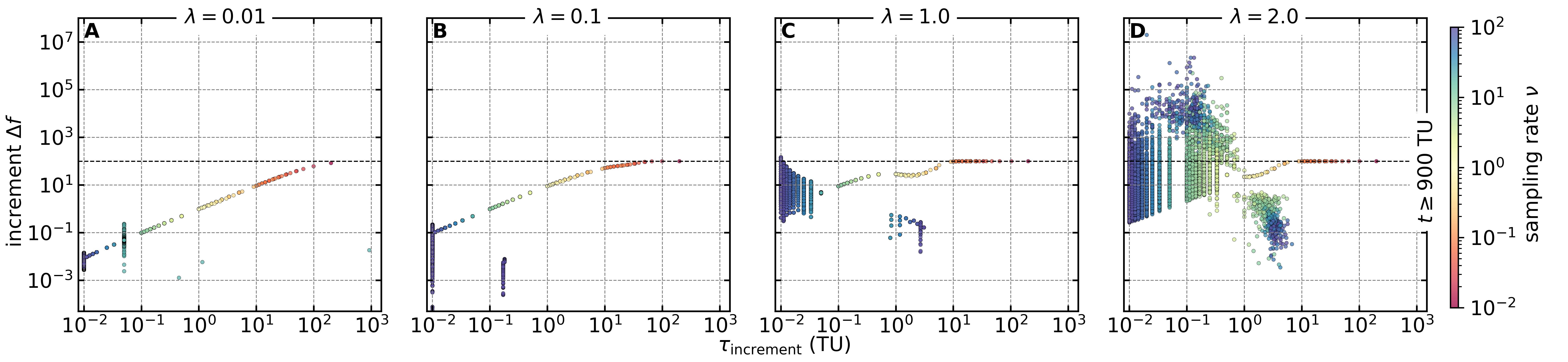}
\caption{{\bf A nonlinear relationship between time intervals and intensities of feedback surges in homeostasis.} {\bf (A--D)} show scatter plots of feedback surges space for the homeostatic regime, $t \geq 900$ TU, namely, in y-axis the increment effects and in x-axis the interval between these increments. The dots' color indicates the sampling rates $\nu$ corresponding to the color code located to the right of the figure. Decaying rates $\lambda$ for each graph are displayed at the column top. All axes are in logarithmic scale, and the rates and time are relative to $\rho$.} \label{fig:tauf_vs_dosef_bar_time}
\end{figure}

The scatter plots in Figure~\ref{fig:tauf_vs_dosef_bar_time} depict the space of feedback surges. Graphs {\bf (A--D)} display overlapping data points representing dynamics at homeostatic regime for $t \geq 900$ TU. They were computed for a wide range of sampling rates $\nu$, equally spaced in logarithm scale from $10^{-2}$ to $10^{2}$.

For decaying rate $\lambda = 0.01$, graph {\bf (A)}, the feedback surges agenda forms a straight line exhibiting a wide range of linear dependence between increment $\Delta f$ and $\tau_{\rm increment}$. Note that agendas with lower sampling rates, $\nu < 1$ (reddish dots), exhibit a more regular pattern. However, for $\nu \geq 1$ (greenish to bluish dots), the increments become smaller and some variability begins to appear. For example, when $\nu \geq 10$, some agendas which increments vary up to 3 orders of magnitude. As $\lambda$ increases, graphs {\bf (B)} to {\bf (D)}, the linear relation between $\Delta f$ and $\tau_{\rm increment}$ becomes a non-monotonic curve. For $\nu < 1$ (yellowish to reddish dots) the increment intensities are fixed at 90. For higher $\nu$, the heterogeneity in the agenda increases. Homeostasis is established by increments with: {\it i.} small increments and intermediate $\tau_{\rm increment}$, {\it ii.} large increments and low $\tau_{\rm increment}$, and {\it iii.} highly variable increments with low $\tau_{\rm increment}$.

\subsection{The dynamics of the probability for the promoter to be ON is confined within a small domain during homeostasis}

\begin{figure}[!h]
\centering
\includegraphics[width=1\linewidth]{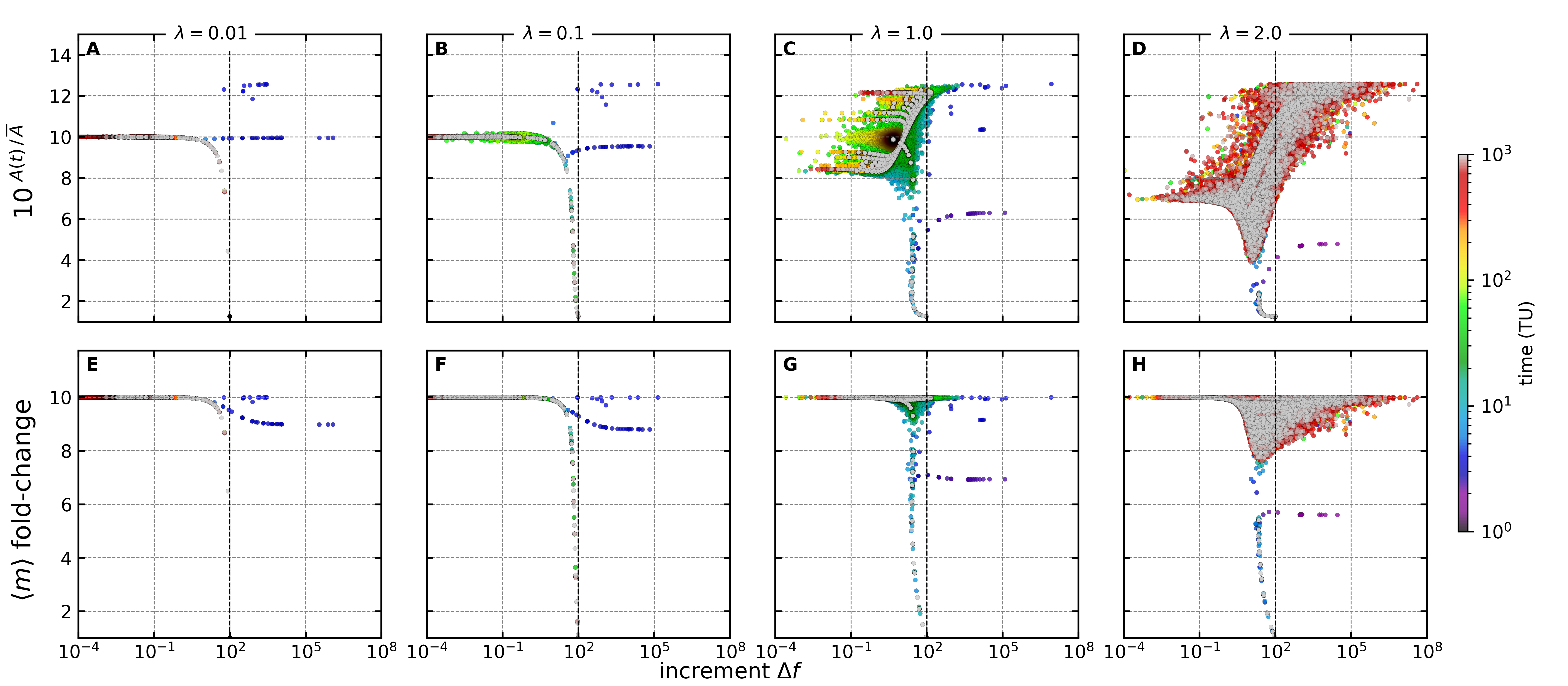}
\caption{{\bf Probability for the promoter to be ON and respective average RNA fold-change during homeostasis.} {\bf (A--D)} and {\bf (E--H)}, depict, respectively, the time course of the dependence between increment intensities and the ON state probability, $A(t)$ or the respective fold change in the average RNA levels, $\langle m \rangle\,(t)$, for a comprehensive range of sampling rate $\nu$. The marker colors represent time in TU (see the color code on the right). The gray points indicate the homeostatic regime. The horizontal axes are displayed in logarithmic scale, while the vertical ones denote both the normalized ON-state probability and the fold-change in $\langle m \rangle$. The normalized probabilities of the ON-state are better visualized through rescaling to powers of 10.} \label{fig:a_norm_vs_dosef_bar_time}
\end{figure}

Figure \ref{fig:a_norm_vs_dosef_bar_time} depicts the relation of the feedback surges effect on $f$ to: the gene promoter ON state probability, graphs {\bf (A--D)}; the average fold change in RNA levels, graphs {\bf (E--H)}. We consider a wide range of sampling rates $\nu$. The feedback surges are computed using Eq. \ref{eq:f_update} while the ON probability $A(t)$ and $\langle m \rangle\,(t)$ levels are determined by Eqs.~\ref{eq:probON_t} and \ref{eq:avrg}, respectively. Here, the aimed ON probability $\oa$ for a 10-fold increase in RNA levels is 0.9. It is worth noting that ${10\,}^{A(t)/\oa} = 10$ implies $A(t) = \oa$. The first increment always has an intensity of 90 such that $10^{A/\oa} = 1.26$.

When $\lambda \leq 0.1$, in graphs {\bf (A--B)} and {\bf (E--F)}, the early increments (bluish dots) exhibit intermediate and high intensities, with $A \geq \oa$ and $\langle m \rangle$ fold-change varying between 8 and 10-fold. During the transient (greenish and reddish) and homeostasis (gray) regimes, low increment intensities occur keeping $\langle m \rangle \sim \om$ and $A(t) \sim \oa$. Intermediate increments occur when $A(t)$ and $\langle m \rangle$ are far from the aimed value. For $\lambda = 0.1$, increments in the transient regime fluctuate around $\oa$. In homeostasis, $A(t)$ and $\langle m \rangle$ become more heterogeneous as $\lambda$ increases, accessing more values below the aimed one -- compare graphs {\bf (A, E)} and {\bf (B, F)}.

For $\lambda = 1$ (graphs {\bf (C)} and {\bf (G)}), the increment intensities shift to the intermediate region between $10^{-1}$ and $10^{2}$. Early increments maintain the intensity, however they may occur less effectively, with $A(t) \ll 1$ and $\langle m \rangle$ fold-change around 7. Two behaviors of feedback surges are revealed in the time course: {\it i.} the increment intensities concentrate around 10 with $A(t)$ and $\langle m \rangle$ close to aimed value, {\it ii.} the intensities and ON probability ranging between two states maintain RNA levels around $\om$, namely, low intensity around $10^{-1}$ and ${10\,}^{A(t)/\oa} \approx 8$ ($A(t) = 0.82$), and the state with intermediate intensity around $10^{2}$ and ${10\,}^{A(t)/\oa} \approx 12$ ($A(t) = 0.98$).

For the changing of $\lambda$ value from 1 to 2, graphs {\bf (D)} and {\bf (H)}, the behavior of feedback surges observed in aforementioned case {\it i.} is changed with the attractiveness to a central point during homeostasis being transformed towards case {\it ii.}. However, in this scenario, $\langle m \rangle$ fold-change is less effective and a new minimum appears at ${10\,}^{A(t)/\oa} \approx 4$ with an increment intensity $\sim 10^1$ (see {\bf (D)}). Note that in homeostatic regime, both the range of increment intensity and the distance between the probabilities of two states (case {\it ii.}) increase, RNAs levels may be lower, $\sim 8$-fold, and the feedback surges become more heterogeneous: lower increments vary between $10^{-2}$ to $10^{1}$ while higher increments reach up to $\sim 10^{6}$.

\section{Discussions}\label{sec:discussions}

The feedback-based control model for regulation of a MRG enables one to investigate the conditions under which homeostasis occurs when multiple stochastic processes with distinctive time scales are coupled \cite{Purvis2013,Levine2013}. This effectiveness depends on four timescales governing dynamics: the RNA degradation rate $\rho$, the gene switching frequency $\epsilon$, the decaying rate of the feedback effect $\lambda$, and the sampling rate $\nu$. The 1st and 2nd govern the insulated two-state stochastic model for gene regulation, the 3rd depends on the mechanisms of the feedback control while the 4th can be determined in terms of the sensors which are being affected by the products of the MRG \cite{Filo2023}. Furthermore, our approach provides a framework to investigate cellular phenotype reprogramming, a key goal underpinning cancer therapies \cite{Yesilkanal2021,Yesilkanal2021a}. Indeed, we may see that a gene having parameters set towards a given homeostatic gene expression regime, may be redirected towards an alternative one (Fig. \ref{fig:fold_m_cumulative_vs_t}B and F) accordingly with the relation between the rates $\nu$ and $\lambda$. This result maybe useful to understand the emergence of diseases which are based on the rebalancing the gene expression levels of MRGs. The feedback mechanism prevents one to build an intervention agenda as reported in a previous study \cite{Giovanini2022} though the analysis of the proper reaction rates for fine tuning the control to modulate the expression of the MRG towards a specific aim remains necessary. However, because one may use the exact solutions for the average numbers of RNAs as a replacement to SSA to reach homeostatic dynamics (Fig. \ref{fig:fold_m_cumulative_vs_t}), the exploration of the parameter space is facilitated and prone to application of optimization techniques such as simulated annealing \cite{Kirkpatrick1983,Chu1999}.

The effectiveness of the feedback-based control on providing homeostatic dynamics has a strong dependence on the sampling rate. Indeed, larger values of the sampling rate, {\it e.g.} $\nu = 100$, late feedback surges happen at small deviations from the aimed average number of RNAs (Fig. \ref{fig:fold_m_cumulative_vs_t}D). At this limit, tiny changes in $f_t$ are sufficient for correcting the trajectories for all values of $\lambda$ (Fig. \ref{fig:fold_m_cumulative_vs_t}D). The earlier feedback surges cause larger increments and responses which overshoot $\om$ (Fig. \ref{fig:fold_m_cumulative_vs_t}D). The variability of the feedback surges parameters is increased, see CV in Fig. \ref{fig:m_fold_dosef_tauf_vs_fsampling_bar_cv}B and C. Note that the increments in $f_t$ and interval between feedback surges approach a minimum value as we increase $\nu$ which depends non-trivially on $\lambda$ (Fig. \ref{fig:m_fold_dosef_tauf_vs_fsampling_bar_cv}B and C). For $\lambda = 2$, only the minimum in the average increment is observed, breaking the trade-off between the averages of the increment intensity and interval, which occurs at lower $\lambda$.

As the sampling rate is reduced towards the degradation rate of the RNAs, the role of $\lambda$, the decaying rate of the effect of the feedback control on $f_t$, becomes more prominent. {\it E.g.}, for $\nu \leq 1$, the response in RNA levels to feedback surges is not effective for $\lambda \geq 0.5$ (Figs. \ref{fig:fold_m_cumulative_vs_t}A--B, \ref{fig:m_fold_dosef_tauf_vs_fsampling_bar_cv}A). The sampling rates $\nu \geq 1$ provide homeostatic expression levels at the aimed value. When $\lambda > 0.1$ feedback surges show two behaviors: short spaced in time along with highly variable increment; and widely spaced in time along with fixed size increments (Fig. \ref{fig:tauf_vs_dosef_bar_time}C and D). In these cases, the ON state gene promoter probability $A(t)$ is $\nu$-dependent, as $\nu > 10$, the behavior of $A(t)$ changed from a single stable state value around $\oa$ to oscillating between two states, lower and higher than $\oa$ (Fig. \ref{fig:a_norm_vs_dosef_bar_time}C and D). The dynamics of the control for higher values of $\lambda$ is challenging because of the increasing heterogeneity of $\langle m \rangle (t)$ (Fig. \ref{fig:a_norm_vs_dosef_bar_time}G and H).

Bursty gene expression has been widely recognized as a source of randomness inside the cells \cite{Chubb2006,Suter2011,Larsson2019,LeyesPorello2023,Fukaya2023}. That is in an apparent contrast with the robustness of cellular phenotype determination, and reconciling those two features is a major challenge of the post genomic era. In our picture, a MRG expresses RNA which is sensed by -- {\it i.e.} interact with -- the products of its target genes by either activating or repressing their activity. The net effect is the balance of the quantities of target gene products being activated or repressed, and that is dependent on the amounts of RNAs expressed from the MRG. When the latter is sufficiently large, no feedback surge is induced while the opposite happens for small quantities of the products of the MRG. The sampling rate is then an effective quantity related to the frequency at which the set of products of the target genes sense the RNAs from the MRG. That sensing happens when the products of the target gene are available in the cell and that availability happens in bursts. As we verified either using the dynamics of the averages or SSA, homeostasis is reachable independently of noise, a result that aids in settling the apparent paradox between randomness and robustness of biological systems.

Our theoretical approach is useful for a phenomenological understanding of homeostatic epigenetic control of MRGs and their implications for the design of gene therapies. Those treatment designs involve orchestrating a large number of chemical compounds having a variety of half-lives and coupling affinities. For example, assume a MRG having transcription guided by a promoter with multiple states whose duration is regulated by multiple transcription factors with various affinities to the regulatory sites of the gene. This is exemplified by the HER2 gene in breast cancer which is upregulated (or downregulated) by factors such as TFAP2, Sp1, PbP, YY1, ETS, YB-1, and EGR2 (MYB, FOXP3, GATA3, PEA2, MBP-1, NOTCH, and RBP-JK). These transcription factors bind to the promoter region, and influence the transcriptional activity \cite{Liu2017}. This regulatory process becomes more complex as these factors have different affinities to the regulatory regions of the gene. The use of mathematical models helps to understand effects of the most relevant mechanisms modulating the expression of a gene and how to redirect it with little harm to the cell.

To employ our model in the design of gene therapies one needs to incorporate a more realistic modeling of pharmacokinetics while considering modeling of the feedback surge effects. Typical pharmaceutical agent(s) have parameters such as maximum tolerated concentration, and activity half-life. For instance, Doxorubicin has a half-life of $\sim$ 30 hours \cite{Tacar2013} while Imatinib, a targeted inhibitor, has a half-life of $\sim$ 18 hours \cite{Peng2005}. That highlights the challenge of engineering a feedback system based on how the pharmacokinetics of a specific agent -- encompassing its absorption, distribution, metabolization, and elimination -- affects its effectiveness as a cellular dynamics controlling tool. The use of a more realistic pharmacokinetic model shall help in understanding and quantifying the modulation of the timing, amplitude, and persistence of gene expression alterations, which ultimately determines the expected therapeutic efficacy.

The transcript degradation rate $\rho$ of the MRG model can be explored together with $f$ on the feedback control of the activity of a MRG. Indeed, small RNAs, such as microRNAs (miRNAs) and small interfering RNAs (siRNAs), are components of post-transcriptional regulation \cite{Carthew2009} that may change the rate of RNA degradation, and consequently, a gene expression \cite{Zhu2022}. These mechanisms of post-transcriptional regulation play an important epigenetic role, as they influence gene expression without altering the underlying DNA sequence. Additionally, since these mechanisms are directly connected to gene-editing technologies, manipulation of small RNAs is one possible therapeutic strategy for targeting different diseases \cite{Zhu2022,Kulsoom2025}.

\section*{Funding}
GG thanks CAPES and PPG-Oncologia (88887.699360/2022-00). AFR thanks NIH (R01 OD010936). Authors thank FAPESP (22/00770-0).

\section*{Acknowledgments}
We thank José R. C. Piqueira, Fuad Kassab Jr. and Felipe M. Pait for helpful discussions on control theory; and Roger Chammas for invaluable discussions on cancer biology.

\end{document}